# Cation controlled wetting properties of vermiculite membranes and its potential for fouling resistant oil-water separation

K. Huang[1,2], P. Rowe[3], C. Chi[1,2], V. Sreepal[1,2], T. Bohn[1,2], K.-G. Zhou[1,2,4], Y. Su[1,2,5], E. Prestat[6], P. Balakrishna Pillai[1,2], C. T. Cherian[1,2,7], A. Michaelides[3], R. R. Nair[1,2,8]

[1]National Graphene Institute, University of Manchester, Manchester, M13 9PL, UK.
[2]School of Chemical Engineering and Analytical Science, University of Manchester, Manchester, M13 9PL, UK.
[3]Thomas Young Centre, London Centre for Nanotechnology, and Department of Physics and Astronomy, University College London, Gower Street, London, WC1E 6BT, UK.
[4]Institute of Molecular Plus, Tianjin University, Tianjin, 300072, China.
[5]Department of Materials, Loughborough University, Loughborough, LE11 3TU, UK.
[6] School of Materials, University of Manchester, Manchester M13 9PL, UK.
[7] Department of Physics and Electronics, CHRIST (Deemed to be University), Bangalore 560029, India.
[8]Henry Royce Institute for Advanced Materials, Oxford Road, Manchester M13 9PL, UK.

**The surface free energy is one of the most fundamental properties of solids, hence, manipulating the surface energy and thereby the wetting properties of solids, has tremendous potential for various physical, chemical, biological as well as industrial processes. Typically, this is achieved by either chemical modification or by controlling the hierarchical structures of surfaces. Here we report a phenomenon whereby the wetting properties of vermiculite laminates are controlled by the hydrated cations on the surface and in the interlamellar space. We find that by exploiting this mechanism, vermiculite laminates can be tuned from superhydrophillic to hydrophobic simply by exchanging the cations; hydrophilicity decreases with increasing cation hydration free energy, except for lithium. Lithium, which has a higher hydration free energy than potassium, is found to provide a superhydrophilic surface due to its anomalous hydrated structure at the vermiculite surface. Building on these findings, we demonstrate the potential application of superhydrophilic lithium exchanged vermiculite as a**

**thin coating layer on microfiltration membranes to resist fouling, and thus, we address a major challenge for oil-water separation technology.**

Hydrated ions at charged interfaces are ubiquitous and play a critical role in our daily life ranging from the control of neurological action, protein folding, to clean energy and water production[1-4]. Clays are one of the best examples of such systems[5]. Owing to their unique properties such as high capillarity, degree of swelling[6], and exceptional thermal and chemical stability, they are widely used in sectors ranging from cosmetics, medicine, agriculture, oil and gas industries, environmental remediation and construction engineering[6-8]. Among the many different types of synthetic and naturally occurring clays, vermiculite is a natural clay mineral that has a layered structure with two silica tetrahedral sheets fused to an edge shared octahedral sheet of magnesium, iron or aluminium hydroxides[9]. The stacking of the layers generates regular van der Waals gaps, which are filled with exchangeable hydrated cations[10,11]. Compared to other clays, vermiculite has higher cation exchange capacity, greater layer charge density, and limited swelling[10,12], making it an ideal candidate for developing membranes. Even though it has been an extensively researched material for decades, the microscopic understanding of its wetting properties or its exploitation as a membrane is limited. Previous studies on the wetting properties of clays suggest that the wetting properties are linked to its surface charge, the nature of interlayer cations, interlayer water content, and the surface topography[13-15]. As far as the effect of exchangeable cations is concerned, a few studies on kaolinite, a non-swelling clay mineral, revealed conflicting results ranging from negligible impact to cation-dependent wettability[13,14,16]. The inconclusive nature of the previous wettability studies could be due to the particulate nature of the clays and the difficulty in obtaining a smooth surface for water contact angle measurements. Given the importance of surface charge on the wetting properties of solids[17,18], it is plausible that the adsorption of ions to the surfaces or interfaces will affect the wetting properties of solids[19]. However, previous studies reveal that even though it can influence the wetting properties, a large wetting transition from hydrophilic to hydrophobic has not been observed. In this paper, we successfully demonstrate the wetting transition of vermiculite laminates (V-laminates) from superhydrophilic to hydrophobic by cation exchange technique and also explore their potential application in designing fouling resistant oil-water separation membranes.

The development of superwetting materials that can be used to modify the performance of membranes for ultra/microfiltration has recently gained increased attention because of the potential this holds to reduce membrane fouling, and so address a major challenge in e.g. the area of oil-water separation[20-24]. Most studies reported so far have focused on the surface modification of metallic mesh and textile membranes with superwetting materials. Due to the large pore size of these membranes, this can only be applied for the separation of immiscible oil/water mixtures at lower pressures (mainly under gravity) and is not suitable for emulsions[21,22]. There have also been attempts to modify the surfaces of ultra/microfiltration polymeric membranes by superwetting materials, but due to the instability of the hydration layer upon prolonged exposure to fouling environments, these membranes were found to be unsuitable for continuous filtration[21,22]. Here we show that a simple and scalable coating of superhydrophilic LiV on polymeric microfiltration membranes leads to a robust fouling resistant membrane.

Stable aqueous dispersions of lithium vermiculite (LiV) (inset of Fig. 1a) flakes were prepared from thermally expanded vermiculite crystals using a reflux ion exchange methodology as reported previously[9,11] (Methods). The transmission electron microscopy (TEM) (Fig. 1a) and atomic force microscopy (AFM) image (Supplementary Fig. 1) confirms that the individual flakes are defect free and have a thickness of ≈ 1.5 nm. Micrometre-thick (5 μm) V-laminates (Fig. 1b) were prepared from the above dispersion using vacuum filtration as described in Methods. We used a simple ion exchange method to replace Li ions in the interlayers of the vermiculite membranes to other desired cations such as potassium ($K^+$), calcium ($Ca^{2+}$), lanthanum ($La^{3+}$), and tin ($Sn^{4+}$) (Methods)[10]. Inductively coupled plasma mass spectrometry (ICP-MS) analysis confirms that Li has been completely interchanged with the desired cations (Supplementary Table 1).

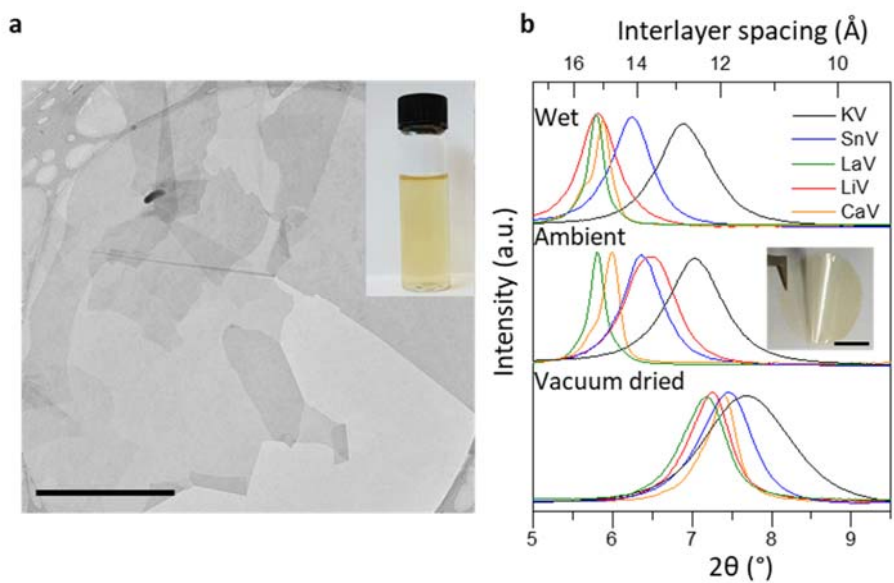

**Fig. 1 | Cation exchanged vermiculite laminates. a**, Transmission electron microscope (TEM) image of exfoliated lithium vermiculite flakes. Scale bar, 10 µm. Inset: Photograph of an aqueous dispersion of exfoliated vermiculite. **b**, X-ray diffraction (XRD) pattern of free-standing K-, Sn-, La-, Li-, and Ca-vermiculite laminates in vacuum dried (12 h), ambient (~ 40% relative humidity) and wet states. Inset: Photo of a 5-µm-thick free-standing Li-vermiculite laminate. Scale bar, 1 cm.

It is well known that the cations in interlayer galleries of the bulk vermiculite crystals are hydrated and that the extent of hydration could depend on the competition of cation hydration energy and the electrostatic interaction between the cation and the layer charge[25]. Consistent with this, vermiculite laminates with different interlayer cations show different Interlayer spacing ($d$-spacing) and different hydration behaviour. Fig. 1b shows X-ray diffraction (XRD) of V-laminates vacuum dried for 12 h, then exposed to air (~ 40% relative humidity) and subsequently soaked in water (Methods). The measured $d$-spacings of vacuum dried samples are in the range of 12.2 Å to 11.5 Å, typical of bulk vermiculite with a single hydration layer[26]. On the other hand, when exposed to ambient air the $d$-spacings increase to 12.6 Å, 13.6 Å, 13.9 Å, 14.7 Å and 15.2 Å for K-vermiculite (KV), Li-vermiculite (LiV), Sn-vermiculite (SnV), Ca-vermiculite (CaV), and La-vermiculite (LaV) laminates, respectively. Further exposure to liquid water only increases the $d$-spacing of LiV (13.6 to 15.2 Å), and no change is observed for CaV, LaV, SnV and KV laminates.

The observed swelling behaviour of V-laminates is consistent with swelling of bulk vermiculite[26,27] and further confirms the exchange of Li cations to the other cations used.

Compared to the bulk vermiculite crystals one of the unique features of the exfoliated vermiculite is its potential for use as a membrane or coating. Hence a thorough understanding of the wetting properties of such films is essential. We have used contact angle measurements to characterize the surface wetting properties of V-laminates. Fig. 2a-e show the water contact angle measurements on cation exchanged V-laminates in dry (ambient) and wet states. We find that cation-exchange modifies the wetting properties of V-laminate in a tunable manner. The water contact angle in air for LiV-laminate is 15°±1° whereas it is 56°±2°, 63°±3°, 75°±2°, and 101°±2° for $K^+$, $Ca^{2+}$, $La^{3+}$, and $Sn^{4+}$ exchanged laminates respectively. It is surprising to note that the LiV-laminate is more hydrophilic compared to all other cation exchanged V-laminates and in the wet state, LiV-laminate is superhydrophilic with zero contact angle while all other V-laminates retained the same wetting behaviour. To further demonstrate the uniqueness of the observed cation controlled superhydrophilic to hydrophobic wetting transition in the V-laminates, we have also performed similar experiments with graphene oxide (GO) membranes (Methods). Cation-modified GO membrane has been shown to offer excellent performance in terms of ion sieving[28]. However, in terms of the sensitivity of the wetting properties of GO to cation modification, we find only a small/moderate sensitivity of ~15 degrees (Supplementary Fig. 2).

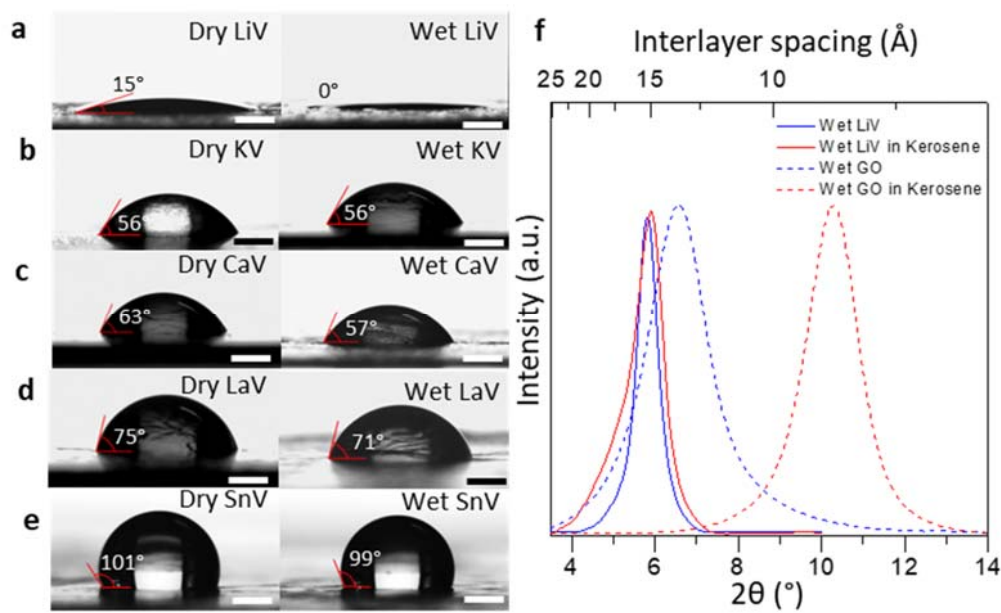

**Fig. 2 | Wetting properties of vermiculite laminates. a-e**, Water contact angle of LiV, KV, CaV, LaV, and SnV-laminates in dry and wet states. Scale bar, 750 μm. We also measured water contact angle for sodium vermiculite (NaV) and obtained a similar value (52°±1°) to that of KV. **f**, Evaluation of water pinning property of free-standing wet LiV laminate measured by capturing XRD from wet LiV laminate before and after immersing in kerosene for a week. As a reference, a hydrophilic GO laminate was also tested in the same experimental conditions, it was found that water molecules were released from the GO membrane after 12 hours of kerosene exposure.

In general, superhydrophilic surfaces retain a hydration layer, and these layers function as a self-cleaning agent by repelling foulants such as biological molecules and oils[24,29,30]. One of the main drawbacks of such surfaces is the instability of the hydration layer upon prolonged exposure to fouling environments[20,24,29]. To investigate the hydration stability of LiV-laminates we have conducted XRD experiments on hydrated LiV-laminates exposed to oil. Fig. 2f shows the stable hydration of LiV-laminate (no change in *d*-spacing) even after a week of kerosene soaking. To rule out the possibility of intercalation of oil molecules into the interlayers and hence providing the same *d*-spacing, we exposed a dry LiV-laminate to kerosene, where we did not observe any intercalation (Supplementary Fig. 3). As a control experiment, we have also conducted similar tests with hydrated graphene oxide (GO) membranes, which have recently been suggested as an antifouling membrane for oil-water separation[31]. As shown in Fig.2f, the *d*-spacing of the wet GO membrane decreases from 13.5 Å to 8.5 Å, when exposed to kerosene for 12 h. This suggests that the interlayer water molecules in the GO membrane have diffused into the oil and the GO membrane has become dry in the oil environment. This in turn implies that the interlayer water in the GO membranes is relatively mobile whereas in the LiV-laminates it is firmly pinned to the interlayer gallery.

The exceptional water pinning features of LiV-laminate suggests its potential for designing antifouling oil-water separation membranes. To study the antifouling properties of LiV, we deposited a thin non-continuous layer of LiV on a hydrophilic microfiltration membrane (polyamide (PA) - nylon) with pore size 1.1 µm (Fig. 3a and Methods) and performed oil-water filtration experiments. The pore size of such membranes was determined by capillary flow porometry (Methods). With increasing the LiV coating thickness, the pore size is observed to decrease and correspondingly the water flux through the membrane also decreases (Supplementary Fig. 4). For further antifouling studies, we optimized the coating thickness to 30 nm such that the wetting time is minimized and the water flux through the coated membrane is comparable to the bare PA membrane (> 75%) (Supplementary Fig. 4).

To test the antifouling properties of the membranes, we have then performed the oil-water separation experiments. Initially, all the membranes were pre-wetted with water to form the surface hydration layer to prevent the oil permeation[29,30]. Such water wetted membranes allow water to permeate but block oil entirely, up to the breakthrough pressure of the membrane (~ 2 Bar for the membranes reported in Fig. 3)[30]. Water flux through such membranes which were continuously in contact with a column of oil (50 ml) was periodically monitored. Fig. 3b shows the water flux as a function of time for the LiV coated PA membrane as well as the reference PA membrane and the GO coated PA membrane. The LiV coated membrane shows a remarkably stable water flux of ~ 7000 Lm$^{-2}$h$^{-1}$ (at 1 bar) even after it is in continuous contact with oil for a week whereas the bare PA suffers severe fouling. The reference hydrophilic GO coating on PA is found to be resistant to fouling for an initial 6 hours (Fig. 3b), but thereafter fouling happens even more severely than for the bare PA. We have also studied the oil repellent property of the LiV coated membrane by measuring the residual organic carbon content in the permeate water and found remarkably high oil repellency (few ppm carbon content as shown in Fig. 3b inset).

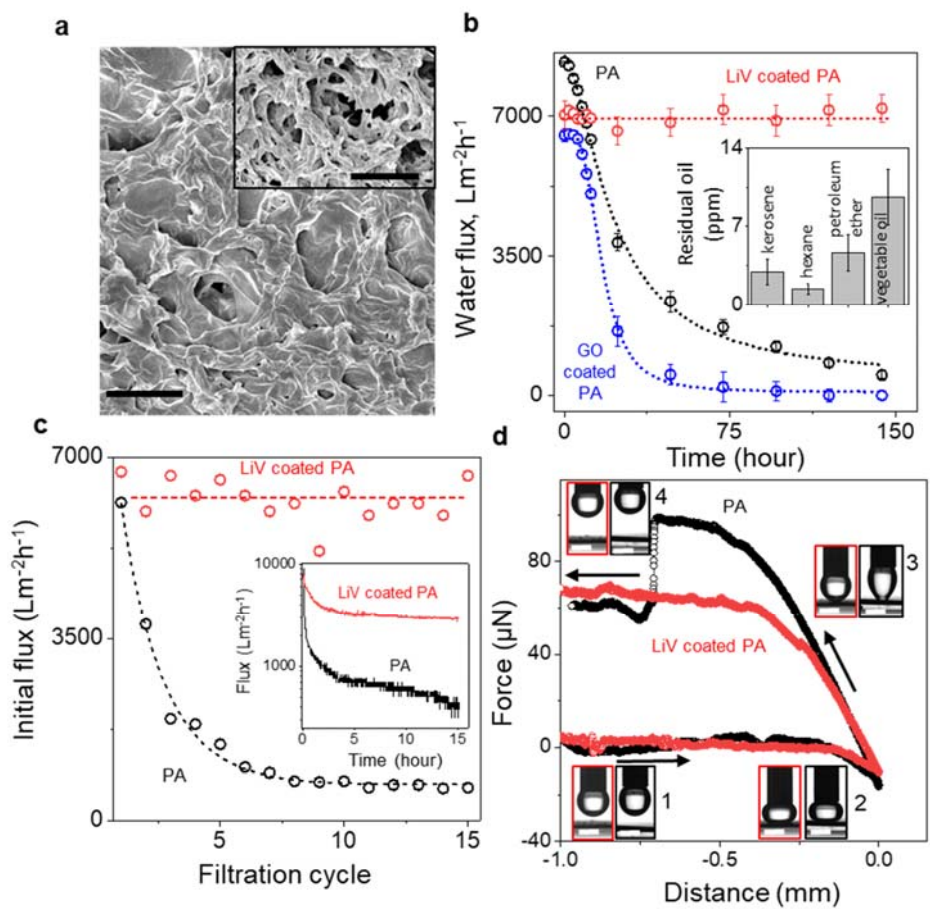

**Fig. 3| Oil-water separation performance. a**, SEM image showing the surface topography of LiV coated (~ 30 nm) PA membrane. Scale bar is 1 μm. Inset: SEM image of bare PA membrane. Scale bar; 3 μm. **b**, Water flux through the LiV coated PA, bare PA, and a reference GO coated (~ 30 nm) PA membrane as a function of time during which membrane was in contact with kerosene. The measurement is performed at 1 bar differential pressure. Error bars denote standard deviations using three different measurements. Inset; residual organic carbon content in the permeate water of the LiV coated PA membrane for different types of oil. Error bars denote standard deviation using measurement at different time intervals. **c**, Initial permeate flux at each filtration cycle during the multiple cycle emulsion separation by dead-end filtration at a pressure of 1 bar. The dotted lines are guides to the eye. Inset; Permeate flux through LiV coated and bare PA during the cross-flow filtration of oil-in-water emulsion at 1 bar with a cross flow velocity of 0.05 m/s. The initial decrease in permeate flux is due to the oil droplet deposition on to the membrane surface, and LiV coating significantly improves this flux decline and provides a ~ seven times higher steady-state flux compared to bare PA. The initial decrease in the flux can be further reduced by increasing the cross-flow velocity (Supplementary Fig. 6). **d**, Force-distance curves recorded while the sample approaches and detaches from the oil droplet (colour coded labels). The adhesion force measurement process involves four major steps: (1) the sample surface approaches the oil droplet, (2) oil contacts the sample surface under a fixed preload, (3) sample

surface leaves the oil droplet leading to deformation of the oil droplet due to oil-sample adhesion force, and (4) sample surface completely detaches from the oil droplet. Arrows indicate the direction of force measurement. Inset; Photographs showing the shape of the oil droplets during the force measurement for LiV coated PA (red outline) and PA (black outline) at the corresponding stages. Scale bar 2 mm.

The observed antifouling properties of LiV coated membranes were found to be robust even for oil in water emulsion separations (Fig. 3c and Methods). Typically, during emulsion separations, oil droplets in the emulsion tend to adhere to the surface of the membrane and hence permanently foul the membranes after multiple filtration experiments[21,22]. Multiple cycles (15 cycles) dead-end filtration data in Fig. 3c shows that the initial water flux (Supplementary Fig. 5) of the LiV coated membrane completely recovers after each cycle of filtration whereas for the reference PA membranes the water flux continuously deteriorated with increasing filtration cycle. This suggests that the adhesion of the oil droplets to the LiV coating is weak, and hence, a simple water rinsing is sufficient to lift the oil droplet from the surface of the membrane. This can be further corroborated from the stable water flux obtained for the LiV coated PA membrane compared to a bare PA membrane in cross-flow filtration (Methods), where a tangential flow of the feed solution with even a smaller velocity of 0.05 m/s detach the oil droplet from the surface and decreases the fouling. The steady-state permeate flux through LiV coated PA membrane is found to be more than seven times higher than the bare PA and the flux increases with increasing cross-flow velocity (Supplementary Fig. 6). To quantify the adhesion force between the oil-droplet and the LiV coating we have measured the dynamic underwater oil-adhesion force (Fig. 3d) and oil droplet roll-off angles (Methods) and found that the adhesion force is below the detection limit (< 1 µN for LiV coated PA and 40 µN for PA ) and the roll-off angle is extremely small (< 3 degrees for LiV and 58.3 degrees for PA) (Supplementary Fig. 7).

To further demonstrate the significance of the LiV coating on the oil-water separation, we have compared the steady-state permeate flux and oil rejection of commercial microfiltration membranes polyethersulfone (PES), and polyvinylidene difluoride (PVDF) in cross-flow filtration before and after LiV coating. Compared to PA, both PES and PVDF were leaky to oil and LiV coating significantly enhances the oil rejection without hugely influencing the flux (Supplementary Fig. 8 and Supplementary Fig. 9) even for less hydrophilic PVDF membranes.

The exceptionally stable antifouling properties of LiV coated membranes could be attributed to the superhydrophilicity of LiV along with its water pinning properties. The superhydrophilic surface of LiV improves the water wetting and the underwater oleophobicity of PA whereas water pinning due to the hydration of Li ions in LiV maintains the membrane in a hydrated state thereby allowing it to retain its oil-repelling characteristics for a long time. To confirm this further, we have also probed the wetting properties of LiV coated PA membranes and found that LiV coating decreases the water wetting time and increases the underwater oleophobicity (Supplementary Fig. 4 and 10).

To test whether the unique superhydrophilicity of LiV is associated with surface charge density, we probed the ionic strength dependence of the contact angle (Methods) and found negligible change with ionic strength, suggesting surface charge is not the crucial factor (Supplementary Fig. 11). To confirm this further, we have also measured the zeta potential of V-laminates, which is directly related to the surface charge density, and its dependence on the ionic strength. Our results show that the zeta-potential varies with ionic strength unlike that of the contact angle behavior (Supplementary Fig. 11).

Further, we estimated the surface free energy from contact angle data using the primary equation of Fowkes' surface energy theory as detailed in Methods. The dispersive component of surface energy is roughly the same for all vermiculite membranes irrespective of the intercalated cation whereas the polar component decreases in the order LiV > KV > CaV > LaV > SnV (Supplementary Fig. 12). Typically the polar component originates from Coulomb interactions between dipoles (e.g. hydrogen bond)[32] and hence this suggests that the origin of the superhydrophilic properties of LiV could be linked to the hydrated structure of the Li cation. To probe this further, we have performed density functional theory (DFT) calculations and ab initio molecular dynamics (AIMD) simulations (Methods). Here we choose to focus on thin water films on the surface of KV and LiV laminates to clearly elucidate the cause of the anomalously hydrophilic nature of the LiV system.

In our analysis, we find that potassium ions exhibit a strong preference for adsorption on top of the hydroxyl groups present on the vermiculite basal plane. In the lithium case, there is a competition between two adsorption sites, one in which the lithium is chelated into the vermiculite siloxane ring and one in which the lithium atom is bound to an oxygen atom adjacent to an aluminium dopant atom, (Fig. 4a and 4b and supplementary Figs. 13 and 14) resulting in a disordered arrangement of lithium cations on the vermiculite surface. This has a strong impact on the ordering of the water contact layer, which we investigate by studying the probability density of finding water oxygen and hydrogen atoms at a given height above the vermiculite surface (Fig. 4a, b), as well as the orientational structuring of water in the first and second contact layers (Fig. 4c, d). The contrast between the highly ordered KV contact layer, and the disordered LiV contact layer can be clearly seen in the plots of the probability density perpendicular to the surface (Fig. 4a, b), where the peaks in the first contact layer of water (approximately 3 Å above the vermiculite surface) in the probability density for water oxygen and hydrogen atoms are well defined and sharp in the case of KV but broad and poorly defined for LiV (below approximately 4.5 Å). Most importantly, we note that unlike the contact layer, subsequent layers of water are far more ordered in the LiV case than in the case of KV, a clear sign of a more hydrophilic surface. The effect of the structuring of the contact layer of water can be further investigated by studying the orientation of water molecules in the two contact layers (Fig. 4c, d). We define two angles, $\theta$ as the angle between the vector normal to the vermiculite basal plane and the O-H bond and $\phi$ as the angle between the plane passing through the water molecule and the plane of the vermiculite surface (Supplementary Fig.15). It can clearly be seen that the structure of the first contact layer in the KV system is strongly distorted from that of the second layer, with a preference for water molecules to orient with one O-H bond directed downwards towards the vermiculite surface. This effect is due to the strong preference of K ions to bind above the vermiculite hydroxyl groups, thereby forcing water molecules to order around them. In the case of the LiV system, the orientations of water molecules in the first and second contact layers are extremely similar, indicating that the LiV surface allows the contact layer of water to orient in a similar way to the more 'bulk-like' layers above. This can be attributed to the large number of available water adsorption sites in the LiV case, as well as the alleviation of steric constraints due to some of the lithium ions adsorbing in the chelated configuration inside the vermiculite surface.

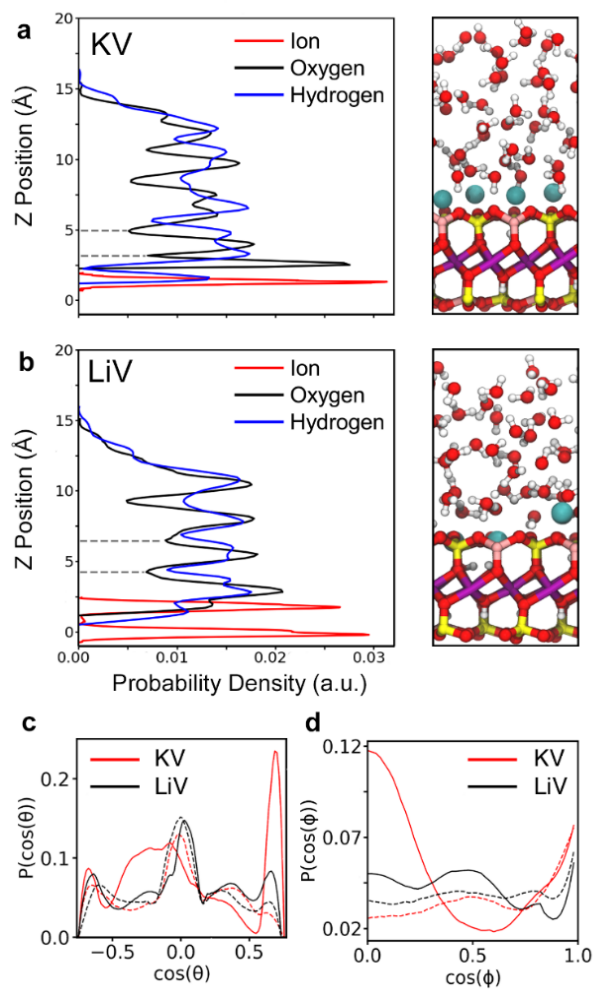

**Fig. 4| Cation hydration structures on vermiculite surface. a-b**, probability density plots for thin water films on vermiculite surfaces substituted with potassium and lithium respectively. The y-axis zero is taken to be the mean z position of the vermiculite surface oxygen atoms. Grey dotted lines indicate the distinction between the first and second contact layers of water. A snapshot is included showing the simulated liquid film and vermiculite structure on the right-hand side. Oxygen atoms are shown in red, hydrogen in white, silicon in yellow, aluminium in pink, magnesium in purple, and lithium and potassium in cyan. **c**, Angular orientation of water molecules in the first and second contact layers of water using the angle between the vermiculite surface normal and water O-H bonds (θ), distributions for the first contact layer are shown with solid lines, dashed lines represent the second layer of water. **d**, Angular orientations of water molecules using the angle between the plane through the water molecule and the basal plane of the vermiculite surface (ϕ), distributions for the first contact layer are shown with solid lines, while the second contact layer is shown with dashed lines.

We propose that the KV surface is made more hydrophobic by the constraints induced by the structuring of water due to the small number of water adsorption sites available. This limited number of adsorption sites causes the interface between KV and water to be highly ordered and thus higher in energy. By comparison, the large number of available adsorption sites at the interface between LiV and water allows the interfacial water layer, and thus subsequent layers, to relax, making the surface more hydrophilic. We further show this by computing the binding energy of individual water molecules in the two contact layers for both systems. We find that in the KV case, the binding energy of water molecules in the contact layer is -0.69 ± 0.09 eV/$H_2O$, while in the lithium case water binds more strongly with an energy of -0.89 ± 0.02 eV/$H_2O$. This effect persists into the second contact layer, where the binding energies are found to be -0.6 ± 0.5 eV/$H_2O$ and -0.7 ± 0.5 eV/$H_2O$. The favorable binding of water molecules in the LiV contact layer explains the observed hydrophilicity and small contact angle of the LiV laminates.

In summary, we have revealed the cation controlled wetting properties of V-laminates, and in particular, we have successfully demonstrated the wetting transition of vermiculite laminates from superhydrophilic to hydrophobic by cation exchange. Our results show that in addition to ion charge and hydration energies, the atomic scale details of the hydration of ions can hugely affect the surface properties of solids and can be utilized for several applications. In addition, by making use of LiV's super-hydrophilicity and stable hydration, we have demonstrated its unique opportunities in designing fouling-resistant oil separation membranes.

**Methods**

**Fabrication of vermiculite laminates:** Vermiculite dispersion was obtained from the thermally expanded vermiculite (Sigma Aldrich, UK) via a two-step ion exchange method as reported previously[9,11]. 50 mg of vermiculite granules were added to 100 mL saturated NaCl (36 wt. %) solution and stirred under refluxing at 100 °C for 24 hours to replace the interlayer cations ($Mg^{2+}$) with $Na^+$. The solution was then filtered out, and the collected vermiculite flakes were repeatedly washed with water and ethanol to remove any residual salt. Sodium exchanged bulk vermiculite was then dispersed in 100 mL of 2M LiCl solution and refluxed for 12 hours followed by filtration and extensive wash with water and ethanol. The filtered product was again dispersed in 100 mL fresh 2M LiCl solution for another 12 hours to secure a complete exchange. Lithium vermiculite (LiV) flakes so-obtained were sonicated in water for 20 minutes in order to exfoliate them into monolayer LiV flakes and subsequently centrifuged at 3000 rpm to remove any multilayers and bulk residues left in the solution. The thickness and the lateral dimensions (~ 1×1μm) of the exfoliated vermiculite flakes (Supplementary Fig. 1) were measured by using Bruker Dimension Fast Scan Atomic Force Microscope (AFM) operating in peak force tapping mode. Electron microscopy characterization of the flakes was carried out by using a FEI Talos F200X transmission electron microscope operating at an acceleration voltage of 200 kV.

The LiV-laminates of thickness ≈ 5 μm were prepared by vacuum filtration of the LiV dispersion through a Whatman Anodisc alumina membrane filter (0.2 μm pore size and a diameter of 25 or 47 mm). The resulting vermiculite films on alumina filters were peeled off from the substrate to obtain free-standing LiV-laminates (Fig. 1). The other cation exchanged vermiculite laminates (V-laminates) were prepared by immersing the LiV-laminate in 1M aqueous chloride solution of the desired cation for an hour.

Graphene oxide (GO) laminates used for the comparison were also prepared by vacuum filtration as reported previously[2]. In brief, the aqueous suspension of GO prepared by bath sonication of graphite oxide flakes (purchased from BGT Materials Limited) were vacuum filtered through Anodisc alumina filters for obtaining free standing GO laminates. Cation-modified GO membranes for the contact angle measurements (Supplementary Fig. 2) were prepared as reported previously[28] by immersing the GO membrane (5 μm) inside 1 molar of the corresponding aqueous chloride solution for one hour followed by rinsing with water.

**Characterization of vermiculite laminates:** We have used inductively coupled plasma atomic emission spectrometry (ICP-AES), and X-ray diffraction (XRD) to characterize the V-laminates. The ICP-AES analysis provided the concentration of the interlayer cations in the V-laminates and confirmed the efficient cation exchange. The samples for ICP-AES analysis were prepared by digesting the V-laminate in a mixture of 1 ml of 38% concentrated HCl, and 1 ml of 70% concentrated $HNO_3$ for overnight. The samples were heated at 70 °C on a hot block in the tubes before being made up to 10 ml with deionized water. The concentration of cations present in the V-laminates was evaluated in moles per mg of the dry V-laminate and is shown in Supplementary Table 1. Cation exchange method used here efficiently replaces the initial cations in the laminates with cations of interest. For example, the concentration of Li-ions in KV-laminate obtained by replacing Li-ions with K-ions is below the detection limit of the instrument (Supplementary Table 1). It should be noted that the $K^+$-ion is present in all the laminates and the starting bulk vermiculite since its strong binding to the vermiculite surface[26,33]. Nevertheless, in KV-laminates the concentration of K-ions is significantly higher than that of other laminates.

XRD experiments were performed to study the laminar structure and swelling properties of the V-laminates with 5 μm in thickness. We used Rigaku smart lab thin film XRD system (Cu-K$\alpha$ radiation) operated at 1.8 kW. For acquiring XRD from dry V-laminates, initially, all the samples were vacuum dried and stored in the glove box for 48 h. The dehydrated samples were sealed in air-tight X-ray sample holder inside the glove box[2,34] and taken out for further XRD measurements. Then the same laminates were exposed to ambient air (~ 40 % RH) for 24 hours, and the measurements were repeated. For wet XRD, the laminates are soaked in the corresponding liquid for a minimum of 30 minutes and then immediately acquired the diffraction data.

**Contact angle measurements:** Sessile drop method (KRÜSS drop shape analyzer, DSA100S) was used to measure the water contact angle on the laminates. Free-standing laminates were placed on a holey flat stage such a way that the central part of the laminate would be on top of the hole.

A micro-syringe needle was used to precisely control the drop volume (2 µl). The needle was lowered slowly until the drop touched the vermiculite laminates and then gently raised. The contact angle measurement module was operated in video mode at a capture speed of 60 frames per second.

**Vermiculite coated PA membranes:** The vermiculite coated PA membranes used for studying the antifouling oil-water/emulsion separation were prepared by filtering the vermiculite dispersion through a porous PA support (170 µm thick EMD Millipore™ PA hydrophilic membrane filters with 47 mm diameter) using a dead-end pressure filtration system (Sterlitech HP4750, 1 bar of overpressure), followed by overnight vacuum drying. The vermiculite coating thicknesses of 15, 30, 45, 60, 75, and 90 nm were fabricated. The coating thicknesses were estimated from the equation,

$$t = \frac{C \times V}{A \times D} \qquad (1)$$

where $t$ is the thickness of the coating, $C$ is the concentration of the vermiculite dispersion, $V$ is the volume of the dispersion deposited, $A$ is the coating area, and $D$ is the density of vermiculite film. The density of the vermiculite film was obtained by measuring the weight and volume of a thick ($\approx$ 15 µm) freestanding V-laminate. It is noteworthy that the estimated thickness of the coating could be slightly overestimated due to the porous structure of the membrane.

The bubble point pore size (largest pore size) of vermiculite coated PA membranes was determined by using flow porometry technique (POROLUX™ 1000)[35,36]. The membranes were fully wetted with a low surface tension liquid (perfluoropolyether/porofil solution) and then sealed inside the sample chamber. Nitrogen gas was allowed to flow into the chamber so as to force the wetting liquid out of the membrane pores. The pressure at which $N_2$ gas overcomes the capillary pressure of the fluid and begins to flow through the wet sample yields the bubble point pore size[35], which is given by,

$$D = \frac{4\gamma \cos\theta}{P} \approx \frac{4\gamma}{P} \qquad (2)$$

where $D$ is the membrane pore size/diameter, and $\gamma$ is the surface tension of the wetting liquid, $P$ is the pressure, and $\theta$ is the contact angle of the liquid.

**Emulsion separation using LiV coated membrane:** The feed oil-in-water emulsion was prepared by dissolving 100 mg of sodium dodecylbenzene sulfonate (SDBS) in 1 L of water followed by mixing with 1 g of kerosene/hexane/petroleum ether/vegetable oil. The mixture was sonicated and stirred for 1 hour to obtain a stable (stable for a week) milky emulsion.

For the dead-end filtration, the emulsion was poured into the pressure vessel fitted with the membrane, and the permeate mass (at a feed pressure of 1 bar) was recorded every 30 seconds for a total of 30 minutes by a computer interfaced electronic scale (Ohaus, Navigator NV). After each 30 min, the membrane was soaked in deionized water to remove the adsorbed oil droplets from the surface of the membrane and repeated the emulsion separation. This process is carried out for fifteen cycles, and the permeate flux is continuously monitored as shown in Supplementary Fig. 5.

The cross-flow emulsion separation was carried out using the Armfield FT17 crossflow filtration system. Before filtering the emulsion, deionized water was filtered through the membrane (1 cm$^2$ area) for 6 hours until a steady water flux was obtained. The emulsion was then poured into the feed tank and filtered through the membrane under a feed pressure of 1 bar and a cross-flow speed of 0.05 m/s to 0.5 m/s (Supplementary Fig. 6). This process was carried out for 15 hours continuously and the permeate mass was recorded by a computer interfaced electronic scale (Ohaus, Navigator NV).

The filtration performance of the membranes for both dead-end and cross flow filtration was characterized in terms of the permeate flux (Fig. 3c and Supplementary Fig. 5) and total organic content in the permeate. During the dead-end filtration, LiV coated PA, and the bare PA membrane has an initial permeate flux of ≈ 6500 Lm$^{-2}$h$^{-1}$, and it decreases by 90% in the first thirty minutes of the filtration process. However, the initial permeate flux got fully recovered for the LiV coated PA membrane in the consecutive cycles of filtration after a simple water rinsing (Supplementary Fig. 5). The flux reduction and its recovery in each cycle of filtration is due to the deposition of the oil droplets on the surface of the membrane and its removal by water soaking respectively. On the other hand, the initial permeate flux continuously declined for the bare PA after each cycle of filtration and reduced to ≈ 10 % of the initial permeate flux after the fifteenth cycle (Fig. 3c).

The amount of oil permeated through the membrane is measured by using the total organic carbon (TOC) content analyser (Shimadzu TOC-VCPN analyser). The separation efficiency of the LiV coated membrane was calculated by oil rejection (*R*) given by

$$R = \left(1 - \frac{C_p}{C_f}\right) \times 100 \qquad (3)$$

Where $C_p$ and $C_f$ are the total carbon content of the collected permeate and the feed emulsion, respectively. Supplementary Fig. 9 shows the oil rejection for the LiV coated PA membrane for different types of emulsions prepared from different oils.

**Wetting properties of LiV coated PA membranes:** KRÜSS drop shape analyzer (DSA100S) was used to carry out all the wetting measurements. The wetting time of the LiV coated PA membrane was studied by recording the time taken to fully spread 2 μL of water dropped on the surface of the membrane[30]. For the precise measurement of wetting time, a high-resolution video camera was operated at a capture speed of 60 frames per second. For example, the wetting time for a 2 μL water droplet added to the LiV coated PA membrane with a pore size of ~ 1 μm (coating thickness of ~ 30 nm) was found to be ~ 0.8 seconds whereas the droplet on bare PA took ~ 1.4 seconds to get it spread completely (Supplementary Fig. 4a). Supplementary Fig. 4b shows the variation in wetting time as a function of LiV coating thickness or the pore size of the LiV coated membrane. It was found that ~ 30 nm is the optimum LiV coating thickness required for the shortest wetting time. The higher coating thickness leads to smaller pore size and lower surface roughness and hence longer wetting time. For coating with thickness ≤ 30 nm, we did not find any significant changes in root mean square (RMS) roughness of the membrane with respect to bare PA membrane (Supplementary Fig. 16). This is mainly due to the flexible nature of the 2D flakes of vermiculites which helps to follow the texture of the underlying PA membrane.

The underwater contact angle measurements were performed as reported previously[37]. A custom-made quartz cell was employed as the water reservoir, where the LiV coated membrane was fixed with coated-side-down on top of the container. The bottom of the cell was connected to a micro-syringe, where the oil droplets were released and floated to the surface of the membrane. The underwater oil contact angle of bare PA was found to be 150°±4 whereas it increases to 168°±3° after coating with 30 nm LiV (Supplementary Fig. 10).

Same setup for the underwater contact angle was employed for measuring the roll-off angle. For the roll-off angle test, an oil droplet (10 µl) was squeezed out from the micro-syringe to the surface of the sample immersed in water. The sample stage was then slowly tilted (10 °/min) until the oil droplet slides off from its original position. The whole process was recorded every 400 ms with the aid of a high-resolution camera, and the roll-off angle is obtained by monitoring the movement of the droplet with respect to its original position (Supplementary Fig. 7).

**Oil adhesion force measurement:** Under-water oil adhesion forces were measured as reported previously[38] using a high sensitivity microelectromechanical balance system (DCAT250, Data Physics). The whole measurement process involves three major steps, namely advancing, contacting and receding. The forces were recorded during the entire step using a high sensitivity microelectromechanical balance. First, an oil droplet (10 µL) was suspended on a metal cap attached to the balance. During the advancing process (step 1 in Fig. 3d), the membrane samples immersed in the water attached to a stage is lifted up slowly at a speed of 0.05 mm/s until the membrane surface contacts the oil droplet. After contacting, a fixed pre-load (10 µN) was applied (step 2 in Fig. 3d). The stage then moves downwards with a speed of 0.02 mm/s (receding, step 3 in Fig. 3d), leading to the detachment of the oil droplet from the sample surface. During this detachment process, at a certain position, the droplet completely detaches from the sample surface indicated by a sudden drop in the force curve (step 4 in Fig.3d). The change in the force during this sudden detachment step was taken as the maximum adhesion force. For bare PA samples, this force is found to be 35.4 ± 3.5 µN. However, for LiV coated PA, no apparent sudden drop in the force was noticed suggesting a negligible adhesion force[38] (below the detection limit of 1 µN). We used 1,2-dichloroethane, hexane, and toluene to replicate the oil for these measurements and did not notice any significant differences between different solvents.

**Influence of ionic strength on contact angle:** To study the impact of surface charge density on the contact angle of V-laminates, we measured the contact angle at different concentrations ($10^{-4}$ M to 1M) of salt solutions. It is known that at high ionic strength, the surface charge density decreases due to the compression of the electric double layer (EDL)[39] and hence studying the contact angle variation with ionic strength directly probe its dependence on surface charge density. These measurements were carried out by investigating the contact angle of different V-laminates using a salt solution of varying concentrations. To avoid any ion exchange during the contact angle measurements, we used the salt solution corresponds to the interlayer cations of the V-laminates (e.g., LiCl solution for LiV-laminate and KCl for KV-Laminates). Supplementary Fig. 11 shows the contact angle as a function of the concentration of salt solution for five different V-laminates. For all five types of V-laminates, contact angle did not show any significant change with the ionic strength. It should also be noted that, for LaV and SnV laminates, the contact angle slightly changes at higher concentration of the salt solution. This is due to the change in pH (salt solution become more acidic) of the salt solution at these concentrations.

**Zeta potential measurements:** To further confirm the absence of correlation between surface charge density and the contact angle, we have measured the zeta potential of different V-laminates and its ionic strength dependence via the streaming potential technique (Anton Paar SurPASS3). These measurements were carried out by placing two V-laminates inside the measuring cell forming a capillary with 100 µm height. Then the test liquid (a mixture of LiCl and KCl) with known ionic strength was injected through the capillary at a specific pressure (200-600 mbar), and the potential difference was measured between the two ends of the streaming channel as the streaming potential.

For samples with a planar surface, its zeta potential can be related with the streaming potential by Helmholtz-Smoluchowski equation[40,41]:

$$\xi = \frac{dU_{str}}{d\Delta p} \times \frac{\eta}{\varepsilon \times \varepsilon_o} \times \kappa \qquad (4)$$

where $U_{str}$ is the measured streaming potential at a specific cross-capillary pressure $\Delta p$, $\kappa$ is the conductivity of the capillary, and $\eta$ and $\varepsilon \times \varepsilon_o$ are the viscosity and dielectric coefficient of the electrolyte solution.

Supplementary Fig. 11b shows the zeta potential for different V-laminates. We found that LiV has the maximum zeta potential and it decreases to nearly zero for LaV and SnV with multivalent ions as the interlayer cations. This could be due to the charge neutralisation of the negatively charged silicate layer with positively charged cations[42]. Even though the variation of zeta potential qualitatively correlates with the change in contact angle (Laminates with higher zeta potential shows minimum contact angle), the large change in the contact angle for LiV-laminates compared to other laminates was not correlating with the zeta potential measurements. This absence of correlation between the zeta potential and the contact angle is further confirmed by measuring the ionic strength dependence of zeta potential. As shown in Supplementary Fig. 11b, the zeta potential decreases with ionic strength as expected, whereas the contact angle is largely independent of ionic strength.

**Surface free energy estimation:** Fowkes' surface energy theory[43,44] describes the contact angle ($\theta$) of a liquid on a solid surface as

$$\frac{\gamma_l(\cos\theta + 1)}{2} = (\gamma_l^d)^{1/2}(\gamma_s^d)^{1/2} + (\gamma_l^p)^{1/2}(\gamma_s^p)^{1/2} \qquad (5)$$

where $\gamma_l$ is surface tension of the liquid, $\gamma_l^d$, $\gamma_l^p$ are the dispersive and polar component of the surface tension of the wetting liquid, respectively, and $\gamma_s^d$, $\gamma_s^p$ are the dispersive and polar component of the solid surface energy, respectively. The surface energy component of the laminates can be determined from eq. 5 by using the contact angle measurements with liquids of known polar and dispersive components[45]. We used the contact angle data of a nonpolar aprotic liquid (diiodomethane[46], $\gamma_l = 50.8$ mN/m, $\gamma_l^p = 0$, and $\gamma_l^d = 50.8$ mN/m) and a polar protic liquid (water[46], $\gamma_l = 72.8$ mN/m, $\gamma_l^p = 51$ mN/m, and $\gamma_l^d = 21.8$ mN/m) on the V-laminates (Supplementary Table S2) and calculated the $\gamma_s^d$ and $\gamma_s^p$ (Supplementary Fig. 12). The Total surface energy ($\gamma_s$) of the V-laminate can be calculated as in eq. 6.

$$\gamma_s = \gamma_s^d + \gamma_s^p \qquad (6)$$

**Density Functional Theory (DFT) Geometry optimisations and system model:** Spin polarised DFT calculations were performed using the CP2K[47] software package using the Quickstep algorithm, with the PBE-D3[48,49] functional, Goedecker-Teter-Hutter pseudopotentials[50,51], a 350 Ry plane-wave cut-off, a relative cut-off of 60 Ry and a DZVP molecularly optimised basis set[52].

The crystal structure of the bare vermiculite surface was optimised by first optimising the unit cell volume of the vermiculite structure, until the total energy is converged to within $10^{-4}$ eV, this was followed by an optimisation of the ionic positions until the total energy was converged to $10^{-4}$ eV. We study a model system in which 25% of the silicon atoms in the basal plane of the Vermiculite structure have been substituted with aluminium, resulting in a net surface charge of -2 e per unit cell, which is representative of that found in natural vermiculite[33]. In each case, we place a charge equalising number of cations on the vermiculite surface.

**Liquid water film AIMD simulations:** Water films on vermiculite surfaces were studied for Li and K using the same structural model as discussed above. Four independent vermiculite surfaces were prepared for each ion in combinations of their vacuum optimised locations. A 2 × 1 supercell of vermiculite was used for AIMD simulations, with lattice parameters in the in-plane direction of 10.62 × 9.20 Å. 30 Å of vacuum was introduced to these above the bare vermiculite surface. The surface was then hydrated with 40 molecules of $D_2O$. Deuterated water is used to lessen the need to consider nuclear quantum effects and to allow for the use of a time step of 0.5 fs. Replica simulations were performed in the NVT ensemble at 300 K, using a coloured noise thermostat[53] and were equilibrated for 5 ps before statistics were collected for analysis. After optimisation and equilibration, there remained approximately 20 Å of vacuum between the surface of the liquid film and the periodic image of the vermiculite surface. In total, 35 ps of simulation time was analysed for the K system, and 40 ps for the lithium system.

**Angular distributions of water molecules:** We employ two metrics to study the angular distribution of water molecules above the vermiculite surface. We denote the angle between the direction of the O-H bonds of the water molecules and the surface normal to the vermiculite surface as θ. We also compute the angle between the vermiculite basal plane and the plane passing through all three atoms of the water molecule, which we denote as ϕ. This is illustrated in Supplementary Fig.15.

**Water Binding energy calculations:** Water binding energies were estimated for water molecules in the first and second contact layers of water. This was done by performing independent DFT geometry optimisations on snapshots selected from the AIMD simulations. 10 snapshots for the first and second layer for both Li and K ions were selected. The binding energy is then defined as

$$E_b = E_{tot} - (N_{wat} * E_{wat} + E_{surf}) \qquad (7)$$

Where $E_{tot}$ is the total energy of the optimised surface with either one or two contact layers of water, $N_{wat}$ is the number of water molecules in the contact layers, $E_{wat}$ is the total energy of the isolated water molecule in vacuum and $E_{surf}$ is the total energy of the optimised vermiculite surface with ions.

**Acknowledgements**

This work was supported by the Royal Society, Engineering and Physical Sciences Research Council, UK (EP/K016946/1), British Council (award reference number 279336045), Graphene Flagship, and European Research Council (contract 679689). We thank Mr. Paul R. Lythgoe at Manchester Analytical Geochemistry Unit, School of Earth and Environmental Sciences, the University of Manchester for assisting with ICP measurements. HAXPES and XRD were performed in the Henry Royce Institute for Advanced Materials, funded through EPSRC grants EP/R00661X/1, EP/P025021/1 and EP/P025498/1. We thank Ben Spencer for helping with HAXPES measurements. A.M. acknowledges support from the European Research Council under the European Union's Seventh Framework Programme (FP/2007-2013) / ERC Grant Agreement No. 616121 (HeteroIce project). We are grateful for computational support from the UK Materials and Molecular Modelling Hub, which is partially funded by EPSRC (EP/P020194), for which access was obtained via the UKCP consortium and funded by EPSRC grant ref EP/P022561/1. The authors acknowledge the use of the facilities at UCL Grace High Performance Computing Facility (Grace@UCL), and associated support services.

# Supplementary Information

**Supplementary Figures**

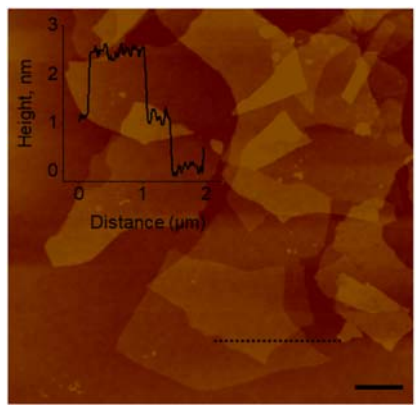

**Supplementary Fig. 1| AFM characterisation.** AFM image of vermiculite flakes drop-casted on a silicon wafer. Inset: Height profile along the dotted line showing vermiculite flakes has an average thickness of ≈1.5 nm. Scale bar, 750 nm.

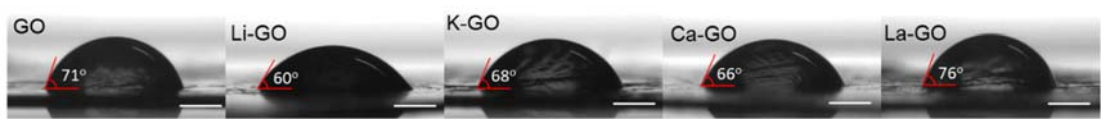

**Supplementary Fig. 2| Contact angle of cation modified graphene oxide (GO) membranes.** Water contact angle of pristine GO and Li, K, Ca, and La modified GO membranes. Scale bar, 750 µm.

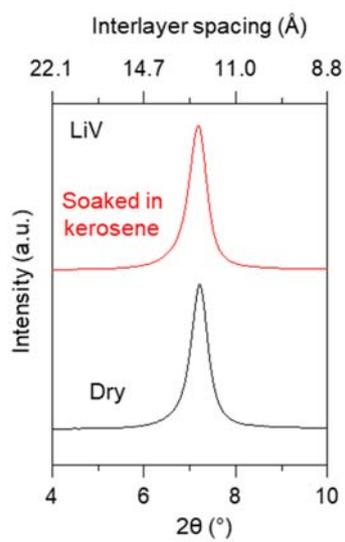

**Supplementary Fig. 3| Non-swelling Vermiculite laminate in oil.** X-ray diffraction (XRD) from a vacuum dried free-standing LiV-laminate and the same membrane soaked in kerosene for 48 h (colour coded labels).

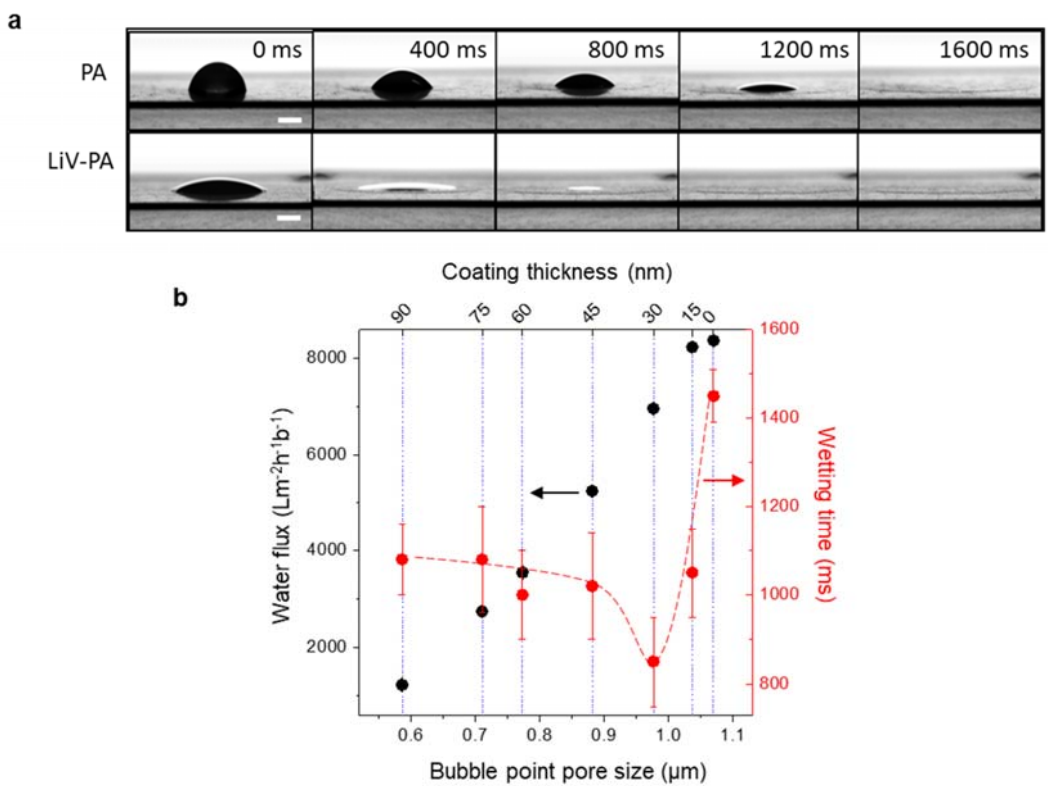

**Supplementary Fig. 4| Water wetting and water flux through LiV coated polyamide (PA) membrane. a**, Water wetting behaviour of the PA and LiV coated (30 nm) PA membranes evaluated by contact angle measurements in dynamic mode. Scale bars; 1000 μm **b**, Water flux and water wetting time for LiV coated PA membranes with various coating thickness or pore size (colour coded axis). The dashed line is a guide to the eye. Error bars denote standard deviations using five different measurements. Water flux was measured by filtering 200 ml of water (after reaching into a steady state flux by filtering water for lmore than one hour) using a dead-end pressure filtration system at a pressure of 1 bar. The high-water flux and low wetting time of 30 nm LiV coated PA membranes makes them as a choice for antifouling studies.

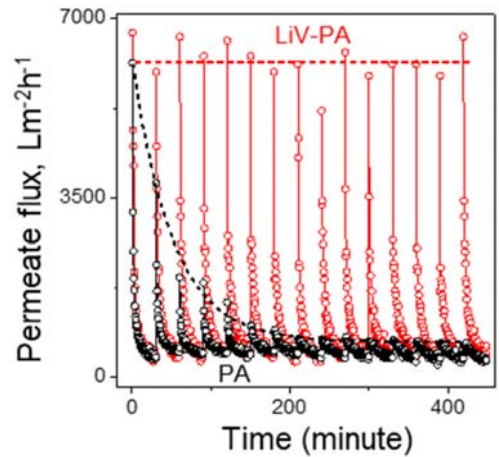

**Supplementary Fig. 5| Cyclic emulsion separation.** Permeate flux as a function of time during the multiple cycle emulsion separation by dead-end filtration at a pressure of 1 bar. The dotted lines are guide to eye for the initial permeate flux at each filtration cycle. The decrease in the permeate flux with time in each cycle is due to the oil droplet deposition on the surface of the membrane. These droplets were easily removed by water rinsing after each cycle in the case of LiV coated membrane whereas it fouls the bare PA membrane permanently.

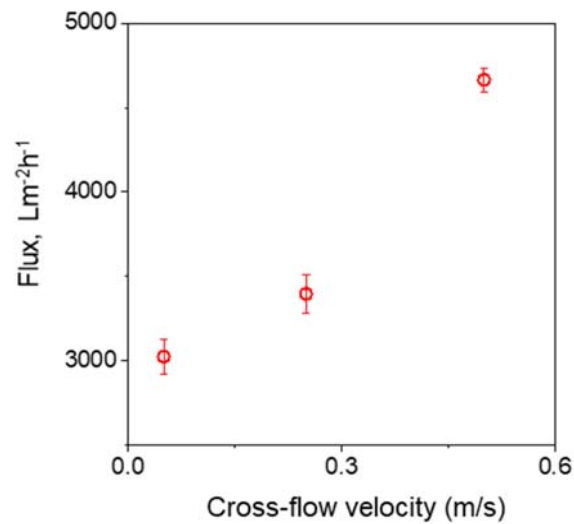

**Supplementary Fig. 6| Cross-flow emulsion separation.** Steady-state permeate flux of LiV coated PA membrane as a function of cross-flow velocity during the emulsion separation at 1 bar pressure.

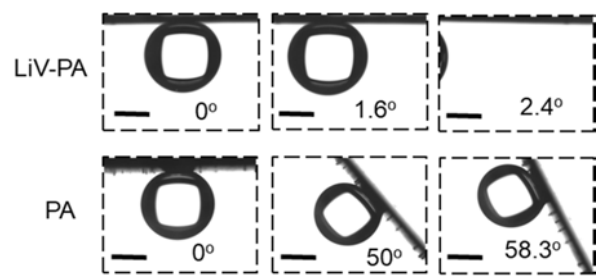

**Supplementary Fig. 7| Oil droplet roll-off angle.** Photographs of an oil droplet (kerosene, 10 µL) on the surface of LiV coated polyamide (LiV-PA) and bare polyamide (PA) membrane at different tilt angles. Scale bars, 1 mm. For the LiV-PA membrane, at 1.6° the droplet starts to slide and completely rolls off at 2.4° whereas, for the bare PA, the droplet starts to slide at 50° and completely rolls at 58.3°. The negligibly small roll-off angle for LiV-PA surface implies a remarkably weaker adhesion force between the oil droplet and the membrane surface.

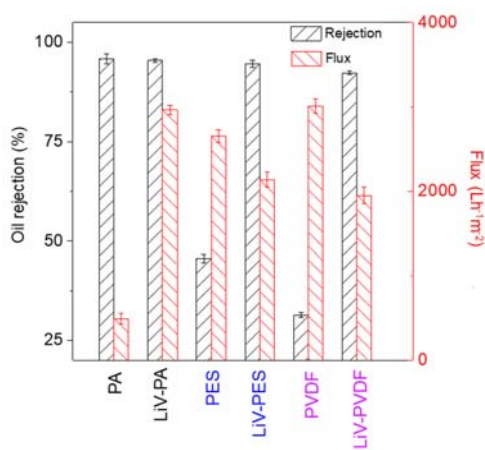

**Supplementary Fig. 8| Emulsion separation performance of LiV coated microfiltration membranes.** Oil rejection and steady state permeate flux for polyamide (PA), polyethersulfone (PES), and polyvinylidene difluoride (PVDF) before and after LiV coating. The enhanced oil rejection for LiV-PES and LiV-PVDF compared to bare PES and PVDF is due to the increase in underwater oleophobicity after LiV coating (Supplementary Fig. 12). The test was carried out with 0.05 m/s cross-flow speed at 1 bar pressure. For PA and PVDF, 30 nm LiV coating was used whereas for PES due to its smooth surface topography, 5 nm coating was sufficient. Increasing the coating thickness leads to a significant reduction in flux. PES and PVDF (PVDFV0.2) membranes with a pore size of 0.22 µm were purchased from Merck Millipore and Sterlitech, respectively.

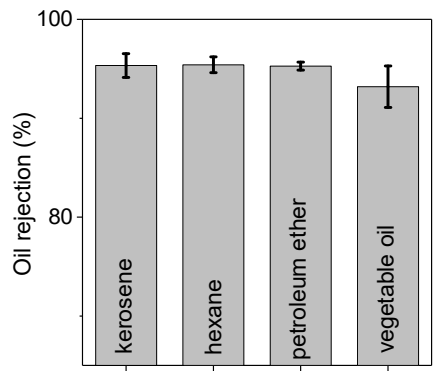

**Supplementary Fig. 9| Emulsion separation.** Oil rejection of the LiV coated PA membrane for different types of emulsion prepared from different oils. Error bars denote standard deviations using three different samples.

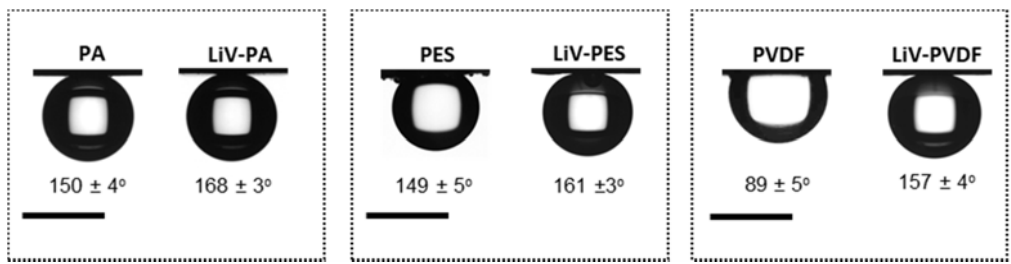

**Supplementary Fig. 10| Underwater oleophobicity.** Underwater oil contact angle measured on bare and LiV coated polyamide (PA), polyethersulfone (PES), and polyvinylidene difluoride (PVDF). Scale bar; 1.5 mm. The coating thickness was 30 nm for PA and PVDF whereas 5 nm for PES.

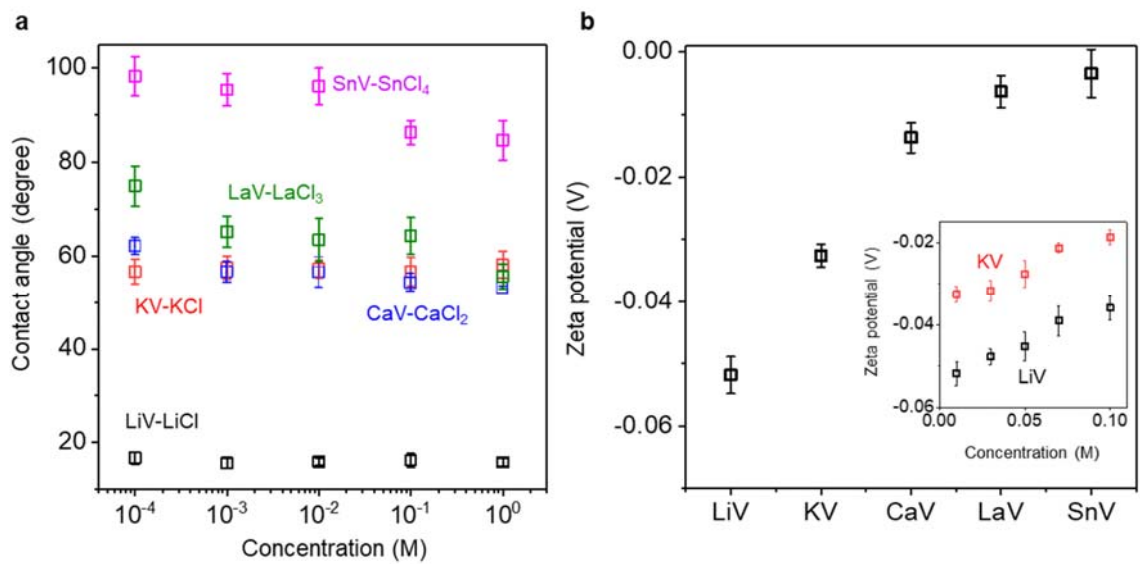

**Supplementary Fig. 11| Effect of ionic strength on contact angle and zeta potential. a**, The contact angle as a function of ionic strength of the solution for various V-laminates. **b**, Zeta potential obtained for various V-laminates. Inset: variation of zeta potential as a function of ionic strength of the solution for KV- and LiV-laminates.

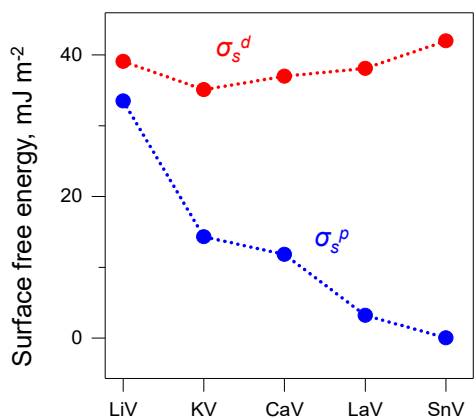

**Supplementary Fig. 12| Surface free energy for different V-laminates.** Dispersion ($\sigma_s^d$) and polar ($\sigma_s^p$) components of the surface free energy for various vermiculite laminates estimated from the contact angle data.

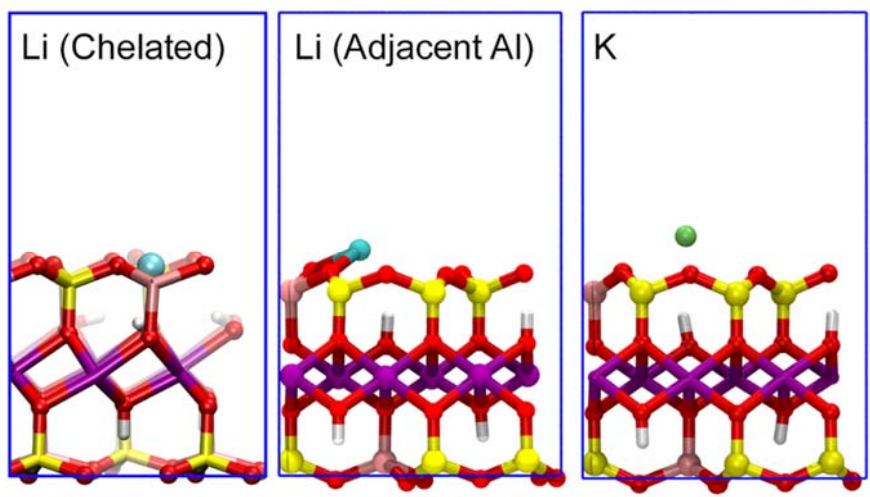

**Supplementary Fig. 13| Ion binding configurations.** Showing two binding configurations for lithium and potassium ions. The chelated configuration for potassium is not found to be stable, this is likely due to the significantly larger ionic radius of the potassium cation. Oxygen atoms are shown in red, hydrogen in white, silicon in yellow, aluminium in pink, magnesium in purple, and lithium and potassium in cyan and green, respectively.

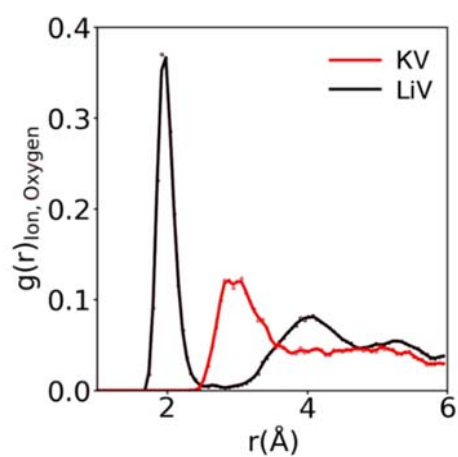

**Supplementary Fig. 14| Ion-Oxygen Radial Distribution Functions.** Radial distribution functions (RDF) averaged over independent trajectories. The first peak in the lithium RDF corresponds to the hydration of the lithium ions bound to oxygen atoms adjacent aluminium dopants. The second, broader peak corresponds to water molecules hydrating the chelated lithium ions.

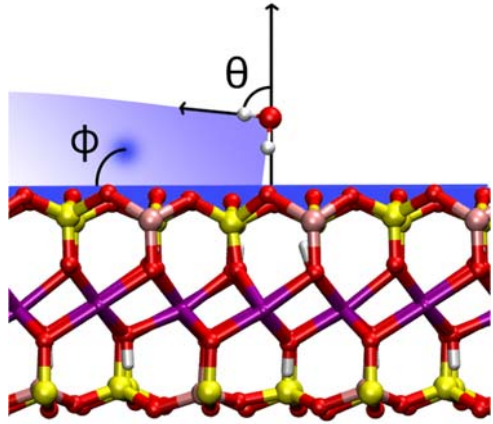

**Supplementary Fig. 15| Angular Definitions.** Illustration of angular definitions for water molecules above the vermiculite surface. θ is defined as the angle between the vermiculite surface normal and water O-H bonds. ϕ is defined as the angle between the plane through the water molecule and the basal plane of the vermiculite surface.

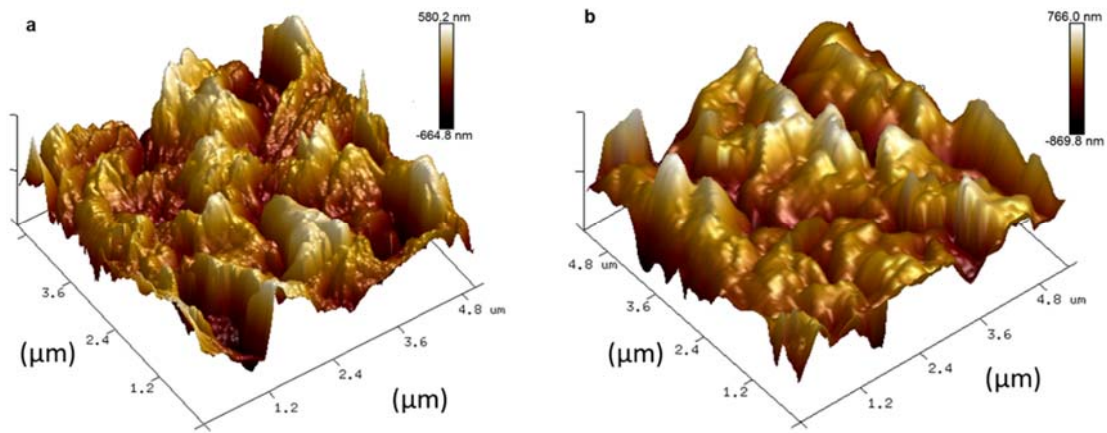

**Supplementary Fig. 16| Root mean square (RMS) roughness of the membranes.** Atomic force microscopy (AFM) image of bare PA (a) and 30 nm LiV-PA (b). AFM Images were captured using a Bruker Dimension FastScan AFM operating in peak force tapping mode. The average RMS roughness obtained from three different samples of bare PA membrane and LiV coated polyamide membrane is 324 ± 112, and 376 ± 58, respectively.

| Moles per mg | LaV | KV | LiV | CaV | SnV | Bulk Vermiculite |
|---|---|---|---|---|---|---|
| K | $1.3\times10^{-7}$ | $8.6\times10^{-7}$ | $2.7\times10^{-7}$ | $7.6\times10^{-8}$ | $1.7\times10^{-7}$ | $9.6\times10^{-7}$ |
| Li | BDL | BDL | $1.4\times10^{-6}$ | BDL | BDL | BDL |
| Ca | BDL | BDL | BDL | $6.8\times10^{-7}$ | BDL | $3.5\times10^{-8}$ |
| La | $4.5\times10^{-7}$ | BDL | BDL | BDL | BDL | BDL |
| Sn | BDL | BDL | BDL | BDL | $6.7\times10^{-7}$ | BDL |

**Supplementary Table 1| ICP Analysis.** Concentration of cations in various vermiculite laminates. BDL denotes 'below detection limit'.

| Laminates | $\theta_{water}$(°) | $\theta_{diiodomethane}$(°) |
|---|---|---|
| LiV | 15 | 41 |
| KV | 56 | 38 |
| CaV | 63 | 45 |
| LaV | 75 | 43 |
| SnV | 101 | 35 |

**Supplementary Table 2|Contact angles.** Water and diiodomethane Contact angle of different V-laminates